\documentclass[conference,a4paper]{IEEEtran}

\usepackage{cite}

\usepackage{amsmath}
\usepackage{amssymb}
\usepackage{multirow}

\usepackage{array}
\usepackage{graphicx}
\usepackage{bm}
\usepackage[normalem]{ulem}

\usepackage{caption}
\newif\iffull
\fulltrue

\newcommand{\eps}{\varepsilon}

\newcommand\blfootnote[1]{%
  \begingroup
  \renewcommand\thefootnote{}\footnote{#1}%
  \addtocounter{footnote}{-1}%
  \endgroup
}

\begin{document}

\title{Spatially Coupled LDPC Codes with Non-uniform Coupling for Improved Decoding Speed\vspace*{-0.6ex}}

\author{\IEEEauthorblockN{Laurent Schmalen
}
\IEEEauthorblockA{
\emph{now with} Karlsruhe Institute of Technology (KIT)\\ Communications Engineering Lab (CEL)\\ (e-mail: \{\texttt{first.last}\}\texttt{@kit.edu})}\vspace*{-2ex}\and\IEEEauthorblockN{Vahid Aref
}
\IEEEauthorblockA{
Nokia Bell Labs, Stuttgart, Germany.\\ (e-mail: \{\texttt{first.last}\}\texttt{@nokia-bell-labs.com})}}
\maketitle

\begin{abstract}
We consider spatially coupled low-density parity-check codes with finite smoothing parameters. A finite smoothing parameter is important for designing practical codes that are decoded using low-complexity windowed decoders.
By optimizing the amount of coupling between spatial positions, we show that we can construct codes with improved decoding speed compared with conventional, uniform smoothing constructions. This leads to a significantly better performance under decoder complexity constraints while keeping the degree distribution regular.
We optimize smoothing configurations using differential evolution and illustrate the performance gains by means of a simulation.
\end{abstract}

\section{Introduction}
\label{sec:intro}
\blfootnote{The work of L. Schmalen and V. Aref has been performed in the framework  of  the  CELTIC  EUREKA  project  SENDATE-TANDEM  (ProjectID  C2015/3-2),  and  it  is  partly  funded  by  the  German  BMBF  (Project  ID16KIS0450K).}
Low-density parity-check (LDPC) codes are widely used due to their outstanding performance under low-complexity belief propagation (BP) decoding.
However, an error probability exceeding that of maximum-a-posteriori (MAP) decoding has to be tolerated with (sub-optimal) low-complexity BP decoding.
A few years ago, it has been empirically observed that the BP performance of
some protograph-based, spatially coupled (SC) LDPC ensembles (also termed \emph{convolutional} LDPC codes) can improve towards the MAP performance of the underlying LDPC ensemble~\cite{Lentmaier-ita09}. Around the same time,
this \emph{threshold saturation} phenomenon has been proven rigorously in \cite{Kudekar-it11,Kudekar-it13} for a newly introduced, \emph{randomly coupled} SC-LDPC ensemble.
 In particular, the BP threshold of that SC-LDPC ensemble tends towards its MAP threshold on any binary memoryless symmetric channel (BMS).

SC-LDPC ensembles are characterized by two parameters: the replication factor $L$, which denotes the number of copies of LDPC codes to be placed along a spatial dimension, and the smoothing parameter $w$. This latter parameter indicates that each edge of the graph is allowed to connect to $w$ neighboring spatial positions (for details, see~\cite{Kudekar-it11} and Sec.~\ref{sec:SC-LDPC}).
The proof of threshold saturation was given in the context of uniform spatial coupling and requires both $L\to\infty$ and $w\to\infty$.
 This poses a serious disadvantage for realizing practical codes, as relatively large structures are required to build efficient codes.

In practice, the main challenges for implementing SC-LDPC codes are the decoding complexity and the rate-loss due to termination. The decoding complexity can be managed by employing windowed decoding (WD)~\cite{Iyengar-wd}, however, the average number of operations needed to decode conventional SC-LDPC codes is still larger as for uncoupled LDPC codes. In high-speed communications, e.g., optical communcations, the number of operations that can be carried out per bit is usually heavily limited~\cite{schmalen2015spatially}. Reducing the decoding complexity is hence an important step towards finding practical SC-LDPC codes. Both the WD complexity and the rate-loss scale almost linearly with $w$ 
suggesting to keep $w$ as small as possible.

It has been shown in~\cite{Kudekar-it11,kudekar2015wave} that the decoding behavior of SC-LDPC codes exhibits a profile that behaves like a ``wave'': it shifts along the spatial dimension with ``a constant speed'' as the BP decoder iterates. 
The WD complexity is also inversely
proportional to the wave propagation speed, which
 has been analytically analyzed and bounded in~\cite{aref2013convergence},\cite{el2016velocity} and used to design irregular degree distributions in~\cite{schmalen2015spatially}.  To decrease the WD complexity and the rate-loss,
 we introduced \emph{non-uniform} coupling 
 in~\cite{SchmalenISIT17} to construct codes with $w=2$ and $w=3$ having good thresholds as well as  
 improved decoding speeds.
 It was recognized earlier in~\cite{SchmalenSCC13,JardelNU} that non-uniform protographs can lead to improved thresholds in some circumstances by sacrificing a one-sided convergence of the chain, which is not problematic when using WD. A particular exponential coupling was used in~\cite{noor2015anytime} to guarantee anytime reliability. Non-uniform coupling has also been successfully used in compressed sensing~\cite{krzakala2012statistical}.

In this paper, we extend non-uniform coupling to randomly coupled SC-LDPC ensembles with $w > 3$. We analyze their performance under message passing using density evolution. We show that properly optimized smoothing profiles can achieve significantly higher decoding speeds than their uniform counterparts using regular variable and check node degrees only. 
The numerically optimized smoothing profiles exhibit some interesting properties that may be used in the future for more targeted code design. We verify the properties of the obtained smoothing schemes by means of simulation.

\section{Spatially Coupled LDPC Codes}
\label{sec:SC-LDPC}
We briefly describe the construction of
non-uniformly coupled LDPC codes: we focus on the \textit{random ensemble} which is easier to analyze and exhibits
the general advantages of non-uniform coupling. In general, \emph{protograph-based} ensembles are of more practical interest. All approaches that we present in this paper can be extended to protograph codes, as was shown also in~\cite{SchmalenISIT17}.

\subsection{The Random $(d_v,d_c,\protect\bm{\nu},L,M)$ Ensemble}
\label{sec:randomSC-LDPC}
We now briefly review how to sample a code from a random, non-uniformly coupled ($d_v,d_c,\bm{\nu},L,M$) SC-LDPC ensemble with regular degree distributions.
We first lay out a set of positions indexed from $z=1$ to $L$ on a \emph{spatial dimension}.
At each spatial position (SP) $z$, there are $M$ variable nodes (VNs) and $M\frac{d_v}{d_c}$ check nodes (CNs),
where $M\frac{d_v}{d_c} \in \mathbb{N}$ and $d_v$ and $d_c$ denote the variable and check node degrees, respectively.
The non-uniformly coupled structure is based on the smoothing distribution
$\bm{\nu}=(\nu_0,\dots,\nu_{w-1})$ where $\nu_i>0$,
$\sum_i\nu_i=1$ and $w>1$ denotes the smoothing (coupling) parameter.
The special case of $\nu_i=\frac{1}{w}$ leads to spatial coupling with the uniform smoothing distribution~\cite{Kudekar-it13}.

\begin{figure}[tb!]
\centering
\includegraphics{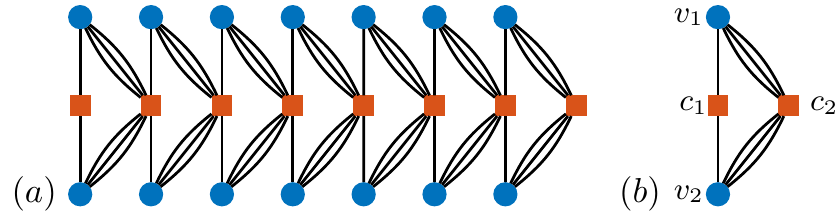}
\vspace*{-1.5ex}
\caption{\label{fig:proto} $(a)$ Protograph of a $(d_v=4,d_c=8,L=7,w=2)$ SC-LDPC ensemble with non-uniform coupling $(b)$ The elementary segment of the chain denoted by the 3-tuple $(4,1,1)$.}\vspace*{-1.5ex}
\end{figure}

For termination, we additionally consider $w-1$ sets of $M\frac{d_v}{d_c}$ CNs in SPs $L+1,\ldots,L+w-1$. Every CN is assigned with $d_c$ ``sockets'' and imposes an even parity constraint on its  neighboring VNs.
Each VN in SP $z$ is connected to $d_v$ CNs in SPs $z,\ldots,z+w-1$ as follows:
For each of the $d_v$ edges of this VN, an SP $z^\prime\in\{z,\ldots,z+w-1\}$ is randomly selected according to
the distribution $\bm{\nu}$, and then the edge is uniformly connected to any free socket of the $Md_v$ sockets arising from the CNs in that SP $z^\prime$.
This graph represents the code with $n=LM$ code bits, distributed over $L$ SPs. Because of additional CNs in SPs $L+1,\ldots,L+w-1$, but also because of potentially unconnected CNs in SPs $1,\ldots,w-1$, the design rate is slightly decreased to $r = 1-\frac{d_v}{d_c}-\frac{1}{L}\Delta$ where
\begin{equation*}
\Delta=\frac{d_v}{d_c}
\left(w\!-\!1\!-\!\sum_{k=0}^{w-2}\left[ \left(\sum_{i=0}^k\nu_i\right)^{d_c} \!\!\!+ \left(\sum_{i=k+1}^{w-1}\nu_i\right)^{d_c}\right]\right),
\end{equation*}
which increases linearly with $w$. 
An exemplary protograph representation of such SC-LDPC ensemble with smoothing distribution
$\bm{\nu}=[\frac{1}{4},\frac{3}{4}]$ 
 is shown in Fig.~\ref{fig:proto}-a). A single 
 element of the chain is separately depicted 
 in Fig.~\ref{fig:proto}-b) exhibiting 
 the non-uniform edge spreading between different 
 SPs.

In the limit of $M$, the asymptotic performance of this ensemble on a binary erasure channel (BEC) can be analyzed using density evolution, with\vspace*{-0.5ex}
\begin{equation}
\label{eq:de}
x_{z}^{(t+1)} = \eps\!\left(\!1\!-\!\sum_{i=0}^{w-1}\nu_i\left(1\!-\!\sum_{j=0}^{w-1}\nu_j x_{z+i-j}^{(t)}\right)^{d_c-1}\right)^{d_v-1}
\end{equation}
where $\eps$ denotes the channel erasure probability and $x_z^{(t)}$ the average erasure probability of the outgoing messages from VNs in SP $z$ at iteration $t$. The messages are initialized as $x_z^{(0)} = \eps$, if $z\in[1,L]$ and $x_z^{(0)} = 0$ otherwise. For $\nu_i=\frac{1}{w}$, \eqref{eq:de} becomes the known 
DE equation for  SC-LDPC codes with uniform coupling
~\cite[Eq.~(7)]{Kudekar-it11}.

\subsection{Decoding Speed \& Windowed Decoder Complexity}
\label{sec:speedWD}
The decoding complexity is an important parameter for practical SC-LDPC codes.
Consider the profile of densities $(x_1^{(t)},x_2^{(t)},\dots)$
 in \eqref{eq:de}. It has been shown in~\cite{Kudekar-it11,kudekar2015wave} that the profile behaves like a ``wave'': it shifts along the spatial dimension with ``a constant speed'' as the BP decoder iterates.
The wave-like behaviour enables efficient \textit{sliding windowed decoding}~\cite{Iyengar-wd}: the decoder updates the BP messages of edges lying in a window of $W_{\textrm{D}}$ SPs $I$ times, and then shifts the window one SP forward and repeats. Thus, the decoding complexity scales with $O(W_{\textrm{D}}ILMd_v)$ as there are $2MLd_v$ BP messages and each BP message is updated $W_\textrm{D}I$ times.

The required window size $W_{\textrm{D}}$ is an increasing function of the smoothing factor $w$~\cite{Iyengar-wd} which implies that we should keep $w$ small. The number of iterations $I>\frac{1}{v}$ where $v$ is the speed of the wave. In the continuum limit of the spatial dimension, $v$ is defined as the amount displacement of the profile along the spatial dimension after one iteration. For the discrete case of \eqref{eq:de}, the speed can be estimated by
\begin{equation}
\label{eq:VD}
v\approx v_D=\frac{D}{T_D},
\end{equation}
where $T_D$ in the minimum number of iterations required for the displacement of the profile by more than $D$ SPs, i.e.,
\begin{equation*}
T_D = \min\{T\in\mathbb{N}\mid x_z^{(t+T)}\leq x_{z-D}^{(t)}, \;{\rm for}\;t>0\;\wedge\;z\leq \lfloor L/2\rfloor \}.
\end{equation*}
The approximation of $v$ becomes more precise by choosing larger $D$. We chose $D=20$ in this paper to minimize the influence of boundary effects, even when $w$ is large.

\section{Non-Uniform Coupling for Higher Decoding Speeds}
\label{sec:randomEns}
In this section, we optimize random non-uniformly SC-LDPC ensembles with the goal of constructing coding schemes that have the highest possible decoding speed. In~\cite{SchmalenISIT17}, we have shown  that for $w\in\{2,3\}$,
non-uniform coupling significantly
improves the threshold of regular SC-LDPC ensembles and also slightly increases the decoding speed.
In this paper, we do not constrain $w$ but instead allow for a larger $w$ to assess the potential increase in decoding speed.
In order to keep the exposition simple, we optimize the speed using the BEC. However, all the techniques can be used in a similar way for other binary symmetric channels as well. In what follows, we describe the cost function used to estimate the speed $v(\epsilon)$.

Assume that we have carried out density evolution~\eqref{eq:de} for $\ell$ iterations and that we have obtained a profile $\bm{x}^{(\ell)}=(x_1^{(\ell)},\ldots, x_L^{(\ell)})$ and assume that $x_{L/2}^{(\ell)} > 0$, i.e., we have not yet achieved full convergence (assuming two-sided convergence is possible). We define the wave position $W_P(\ell)$ as
\begin{equation}
W_P(\ell) = A + \frac{\frac{1}{2}x_{L/2}^{(\ell)}-x_{A}^{(\ell)}}{x_{A+1}^{(\ell)}-x_{A}^{(\ell)}},\label{eq:WP}
\end{equation}
where SP $A\in[1,L/2]$ such that $x_{A}^{(\ell)} < \frac{1}{2}x_{L/2}^{(\ell)}$ and $x_{A+1}^{(\ell)} \geq \frac{1}{2}x_{L/2}^{(\ell)}$. 
Now, let $\bar{T}_{20}$ denote the number of iterations until $W_P\approx 20$ and define the cost function $C^{(1)} = \bar{T}_{20} - W_P(\bar{T}_{20})\frac{2}{L}$. Minimizing $C^{(1)}$  is directly related to maximizing the speed
 $v$, which can be approximated as $v(\epsilon)\approx 20/\bar{T}_{20}$.
In practice, we can find $\bar{T}_{20}$ by carrying out DE using a finite number of $\ell$ iterations, checking whether the code converges and if a wave is currently propagating, followed by finding $A$. We can repeat the procedure and use a binary search over the number of iterations $\ell$ to find $\bar{T}_{20}$ such that $A\approx 20$.

\begin{figure}[t!]
\includegraphics{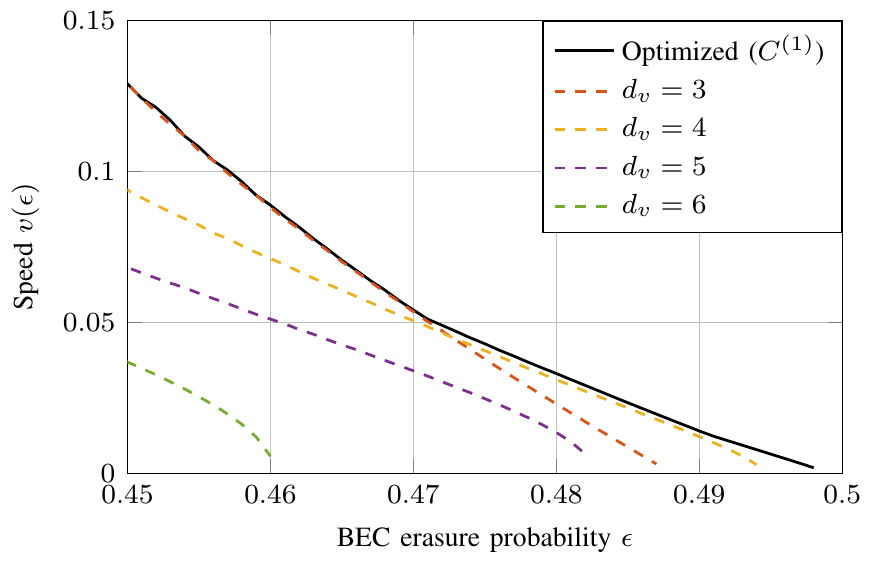}
\vspace*{-3ex}
\caption{Speeds for the $(d_v,2d_v,(\alpha,1-\alpha),L=100)$ ensemble (with $w=2$ and rate $\approx \frac{1}{2}(1\!-\!\frac{1}{L})$) for some regular codes and non-uniformly coupled codes with optimized $\alpha$ and $d_v\in\{3,4,\ldots, 10\}$ obtained using cost function $C^{(1)}$.}
\vspace*{-2ex}
\label{fig:speed_wpopt}
\end{figure}

First, we consider $w=2$ with $\bm{\nu}=(\alpha,1-\alpha)$. This case has a high practical interest as $w=2$ is the smallest smoothing parameter leading to rather small
decoding latency\footnote{This assumes that $M$ is fixed. Increasing $w$ and decreasing $M$ such that $wM$ is fixed would lead to an approximately fixed latency.} and window length $W_{\textrm{D}}$ of WD.
In~\cite{SchmalenISIT17}, we showed that non-uniform coupling improves the BP threshold in this case. In Fig.~\ref{fig:speed_wpopt}, 
the dashed lines illustrate the decoding speeds $v(\epsilon)$ of some regular, uniformly coupled $(d_v,2d_v,(\frac{1}{2},\frac{1}{2}),L=100)$  codes. We can clearly see that the best threshold is obtained for $d_v=4$ (the threshold is given by the largest $\epsilon$ for which $v(\epsilon) > 0$) and the largest speed is shared between $d_v=3$ (for $\epsilon < 0.472$) and $d_v=4$ (for $\epsilon \geq 0.472$). 
We carried out an optimization of the decoding speed with non-uniform coupling: for each $\epsilon$, and each $d_v\in\{3,4,\ldots, 10\}$, we find, using a simple line search, the $\alpha \in[0,\frac{1}{2}]$ (we restrict $\alpha \leq \frac{1}{2}$ to allow only for a left-propagating wave) that minimizes $C^{(1)}$ and compute the corresponding speed $v(\epsilon)\approx 20/\bar{T}_{20}$. For each $\epsilon$, we only keep the value of $d_v$ that minimizes $C^{(1)}$. We can see in Fig.~\ref{fig:speed_wpopt} (solid line) that non-uniform coupling increases significantly the threshold, however, the speed is only marginally improved.

If we increase $w$, the search space for finding a good $\bm{\nu}$ can become too large. In this case, we use differential evolution~\cite{storn1997differential} to find the best smoothing profiles $\bm{\nu}$. The dimensionality of the search space is $D=w-1$ and we use a population consisting of $N_{\mathrm{P}} = 100D$ vectors $\bm{p}_i$, which we initialize by uniformly sampling from the $w$-dimensional polytope of valid $\bm{\nu}$ (as $\sum_i\nu_i=1$, we only need $D=w-1$ values) and keeping those vectors that allow convergence ($\lim_{\ell\to\infty}x_i^{(\ell)}=0$, $\forall i\in\{1,\ldots, L\}$). Inside the differential evolution step, 
for population entry $\bm{p}_i$, we first generate a candidate $\bm{v} = f_{\textrm{Sat}}(\bm{p}_{r_1} + \bm{p}_{r_2} -  \bm{p}_{r_3})$, where $f_{\textrm{Sat}}(\bm{x}) = (\min(|x_1|,1),\ldots, \min(|x_D|,1))$ element-wise restricts the input to the range $[0,1]$ and $(r_1,r_2,r_3)\in\{1,\ldots,N_{\mathrm{P}}\}^3$ such that $r_1\neq r_2$, $r_1\neq r_3$ and $r_2\neq r_3$. We then replace entry $v_j$ ($j\in\{1,\ldots, D\}$) of vector $\bm{v}$ by $p_{i,j}$ with probability $0.33$. If $C^{(1)}(\bm{v}) < C^{(1)}(\bm{p}_{i})$ (i.e., the cost function $C^{(1)}$ for a code where $\nu$ is obtained from the $\bm{v}$, or $\bm{p}_i$, respectively), we store $\bm{v}$ and replace $\bm{p}_i$ in the next iteration with the stored $\bm{v}$. We repeat the steps for all $N_{\textrm{P}}$ entries and iteratively repeat for 1000 iterations. During the evolution, we keep track of the vector leading to the lowest cost $C^{(1)}$.

\begin{figure}[t!]
\includegraphics{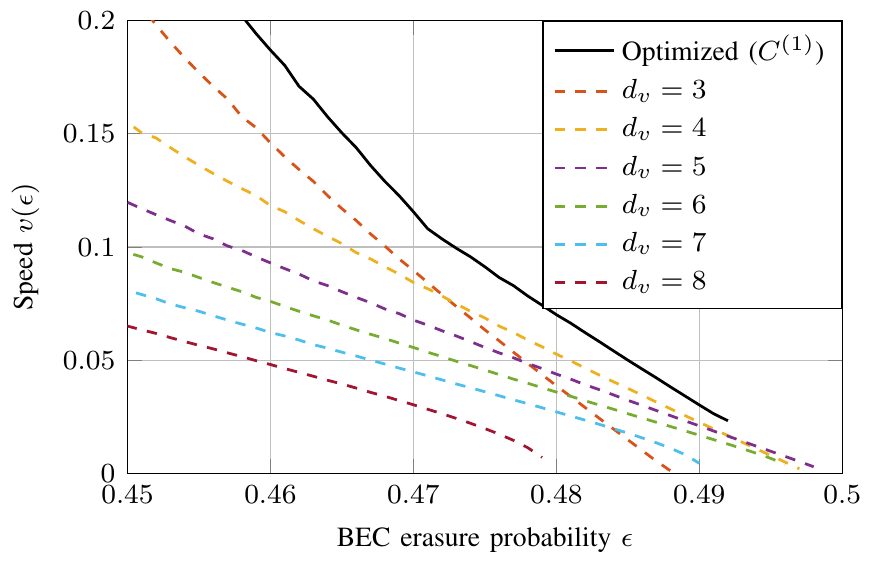}
\vspace*{-3ex}
\caption{Speeds for the $(d_v,2d_v,\bm{\nu},L=100)$ ensemble (with $w=3$ and rate $\approx \frac{1}{2}(1-\frac{1}{L})$) for some regular codes and non-uniformly coupled codes with optimized $\bm{\nu}$ and $d_v\in\{3,4,\ldots, 10\}$ obtained using cost function $C^{(1)}$.}\vspace*{-1ex}
\label{fig:speed_wpopt_w3}
\end{figure}

\begin{figure}[t!]
\includegraphics{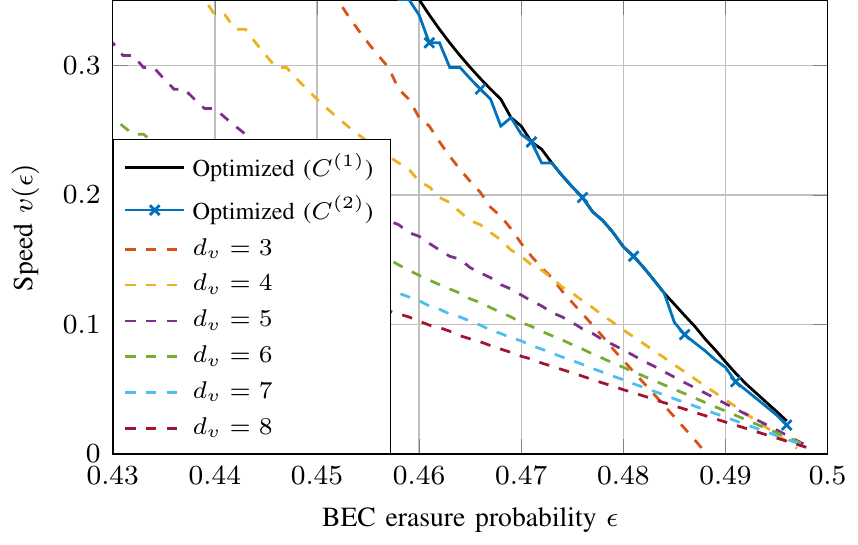}
\vspace*{-1ex}
\caption{Speeds for the $(d_v,2d_v,\bm{\nu},L=100)$ ensemble (with $w=5$ and rate $\approx \frac{1}{2}(1-\frac{1}{L})$) for some regular codes and non-uniformly coupled codes with optimized $\bm{\nu}$ and $d_v\in\{3,4,\ldots, 10\}$ obtained using cost function $C^{(1)}$.}
\vspace*{-3ex}
\label{fig:speed_wpopt_w5}
\end{figure}

\begin{table*}
\caption{Parameters of Codes Selected for Simulation}\label{tab:codeselection}
\centering
\begin{tabular}{ccccccccccccc}
\hline
Code & $w$ & $d_v$ & $\epsilon$ & $\nu_1$ & $\nu_2$ & $\nu_3$ & $\nu_4$ & $\nu_5$ & $\nu_6$ & $\nu_7$ & $\nu_8$ & Rate $R$ ($L=100$) \\
\hline
$\textrm{NU}_{3,\textrm{A}}$ & 3 & 3 & 0.46 & 0.37124 & 0.00835 & 0.62041& --- & --- & --- & --- & --- &  0.49062 \\
$\textrm{NU}_{4,\textrm{A}}$ & 4 & 3 & 0.46 & 0.44135 & 0 & 0.00814 & 0.55051& --- & --- & --- & --- &  0.48556 \\
$\textrm{NU}_{5,\textrm{A}}$ & 5 & 4 & 0.46 & 0.37919 & 0 & 0 & 0.00125 &  0.61956 & --- & --- & --- &  0.48045 \\
$\textrm{NU}_{6,\textrm{A}}$ & 6 & 4 & 0.46 & 0.36720 & 0.00274 & 0.00134 & 0.00091&  0.00098 & 0.62683 & --- & --- & 0.47562 \\
$\textrm{NU}_{8,\textrm{A}}$ & 8 & 4 & 0.46 & 0.37520 & 0.00685 & 0.00356 & 0.00194 & 0.00111 & 0.00015 & 0 & 0.61119 & 0.46573 \\
\hline
$\textrm{Ref}_{3,\textrm{A}}$ & 3 & 3 &---& $1/3$ & $1/3$ & $1/3$& --- & --- & --- & --- & --- &  0.49089 \\
$\textrm{Ref}_{4,\textrm{A}}$ & 4 & 3 &---& $1/4$ & $1/4$ & $1/4$& $1/4$ & --- & --- & --- & --- &  0.48694 \\
$\textrm{Ref}_{5,\textrm{A}}$ & 5 & 3 &---& $1/5$ & $1/5$ & $1/5$& $1/5$ & $1/5$ & --- & --- & --- & 0.48313 \\
$\textrm{Ref}_{6,\textrm{A}}$ & 6 & 4 &---& $1/6$ & $1/6$ & $1/6$& $1/6$ & $1/6$ & $1/6$ & --- & --- & 0.47776 \\
$\textrm{Ref}_{8,\textrm{A}}$ & 8 & 4 &---& $1/8$ & $1/8$ & $1/8$& $1/8$ & $1/8$ & $1/8$ & $1/8$ & $1/8$ &  0.46971 \\
\hline
$\textrm{NU}_{3,\textrm{B}}$ & 3 & 4 & 0.47 & 0.34281 & 0.00501& 0.65218 & --- & --- & --- & --- & --- &   0.49034 \\
$\textrm{NU}_{8,\textrm{B}}$ & 8 & 5 & 0.49 & 0.34485 & 0 & 0.00125 & 0.00443 & 0.00655 & 0.00976 & 0.01293 & 0.62023 & 0.46544\\
\hline
$\textrm{Ref}_{3,\textrm{B}}$ & 3 & 4 &---& $1/3$ & $1/3$ & $1/3$& --- & --- & --- & --- & --- &  0.49039 \\
$\textrm{Ref}_{8,\textrm{B}}$ & 8 & 4 &---& $1/8$ & $1/8$ & $1/8$& $1/8$ & $1/8$ & $1/8$ & $1/8$ & $1/8$ & 0.46971 \\
\hline
\end{tabular}\vspace*{-1ex}
\end{table*}

In Fig.~\ref{fig:speed_wpopt_w3} and~\ref{fig:speed_wpopt_w5}, we compare the speeds of the smoothing profile optimized with cost function $C^{(1)}$ for $w=3$ and $w=5$. In contrast to the case $w=2$, we see a clear advantage in terms of decoding speed $v(\epsilon)$, and less so in terms of decoding threshold (which is already very good for the reference schemes). This advantage gets larger when $w$ becomes larger as well with a very steep decline observed. Again, we see a similar behavior for the reference schemes where the largest speed is obtained for $d_v\in\{3,4\}$. As the evaluation of the cost function becomes increasingly more complex when close to the threshold, we didn't extend the curves towards the threshold (which is similar then for the uniformly coupled case).

In practice, we can often use a simpler cost function, as the binary search to find $\bar{T}_{20}$ can be computationally demanding (especially if carrying out a numerical optimization like differential evolution, requiring a large number of cost function evaluations). Let $\tilde{T}_{10}$ denote the iteration at which SP $z=10$ becomes virtually error free, i.e.,
\[
\tilde{T}_{10} = \min\{t\in\mathbb{N}\ |\ x_{10}^{(t)} < 10^{-19}\}
\]
assuming that the code converges, i.e., $\lim_{\ell\to\infty}x_{i}^{(\ell)}=0$, $\forall i \in\{1,\ldots, L\}$. The value $10^{-19}$ has been chosen as a compromise between accuracy and numerical stability.
 The value of $\tilde{T}_{10}$ hence denotes the number of iterations until the decoding wave fully passes beyond SP $z=10$. 
The cost function $C^{(2)}$ is then given as $C^{(2)} = \tilde{T}_{10} - W_P(\tilde{T}_{10})\frac{2}{L}$, with $W_P(\cdot)$ as defined in ~\eqref{eq:WP}. While the cost function $C^{(2)}$ is easier to calculate, it may be misleading, as due to numerical inconsistencies, $W_P(\tilde{T}_{10})$ may not indicate the true position of the wave. However, it only requires a single execution of DE. To highlight the fact that $C^{(2)}$ can be used as well to optimize smoothing profiles, we carried out  differential evolution for $w=5$, with the same parameters, but using cost function $C^{(2)}$ instead of $C^{(1)}$. For the smoothing profiles $\bm{\nu}$ obtained from the optimization routine with cost function $C^{(2)}$, we compute the value $\bar{T}_{20}$ and the speed $v(\epsilon) \approx 20/\bar{T}_{20}$, which we plot as an additional curve in Fig.~\ref{fig:speed_wpopt_w5} (solid line with markers). We can see that the difference is marginal and the obtained speeds even coincide in some regions. We observe similar results for other values of $w$ (not shown here). Hence, we may use $C^{(2)}$ during optimization as it is significantly less complex to evaluate.
Further increasing $w$ will result in even larger speed gains, as we will also see in the simulation example below.

\section{Simulation Results}

In order to assess the performance of the optimized non-uniformly coupled codes, we carry out a simulation example using the BI-AWGN channel. As SC-LDPC codes are (asmyptotically) universal~\cite{Kudekar-it13}, we use the BEC-based optimization as a simple proxy to find parameters of the codes. Note that the universality has been proven in~\cite{Kudekar-it13} only for the uniformly coupled case and may not apply to the non-uniformly coupled codes. We have seen before that the decoding speed depends on the erasure probability $\epsilon$ and (slightly) different $\bm{\nu}$ are obtained for each different $\epsilon$. We have therefore carried out the optimization for $\epsilon \in\{0.46, 0.47, 0.48, 0.49\}$, yielding an optimized $\bm{\nu}$ and a corresponding $d_v$ for each $\epsilon$. In a next step, we have constructed codes of rate $\approx \frac{1}{2}$ with $d_c=2d_v$ and $M=8000$ bits per SP for each of these four cases.  We have carried out finite length simulations using WD~\cite{Iyengar-wd}  with window length $W_{\textrm{D}}$ and $I$ iterations per decoding window step. We consider two decoder configurations in this paper: In decoder configuration A, we carry out a single iteration ($I=1$) per decoding window, as is frequently done in high speed (e.g., optical) communications. To avoid issues with too short windows (see~\cite{klaiber2018avoiding} for details), we select $W_{\textrm{D}} = 5\cdot w$. Decoder configuration B targets a higher decoding complexity (and hence also a better performance) and selects $I$ such that $\approx 80$ effective iterations are carried out for each bit. The decoder configurations are summarized in Tab.~\ref{tab:decsetup}.
\begin{table}[b!]
\vspace*{-2ex}
\caption{Decoding Setups}\label{tab:decsetup}
\centering
\begin{tabular}{ccccc}
                   & \multicolumn{2}{c}{Decoder Setup A} & \multicolumn{2}{c}{Decoder Setup B} \\
Coupling width $w$ & $W_{\textrm{D}}$ & $I$ & $W_{\textrm{D}}$ & $I$ \\
\hline
3 & 15 & 1      & 15 & 5 \\
4 & 20 & 1      & 20 & ---\\
5 & 25 & 1      & 25 & ---\\
6 & 30 & 1      & 30 & ---\\
8 & 40 & 1      & 40 & 2\\
\end{tabular}
\end{table}
From the finite length simulation, we select the code performing best (which is associated with a certain $\epsilon$).  We call the respective optimized codes  $\textrm{NU}_{w,b}$, where $b\in\{\mathrm{A},\mathrm{B}\}$ denotes the decoder configuration.    We would like to point out that the optimized $\bm{\nu}$ do not vary significantly if we use cost function $C^{(1)}$ or cost function $C^{(2)}$ (see also Fig.~\ref{fig:speed_wpopt_w5} and the discussion above). Interestingly, it appears that the dominant terms in $\bm{\nu}$ are $\nu_1$ and $\nu_w$ and the remaining terms are negligible.

For each $w$, we constructed additionally reference codes of rate $R\approx 1/2$ with uniform coupling and regular $d_v\in\{3,4,5\}$ with $M=8000$ ($d_c = 2d_v$). We have carried out simulations and selected the best uniformly coupled reference code for each $w$ and each decoder setup. The parameters of all the codes are summarized in Tab.~\ref{tab:codeselection}.

\begin{figure}
	\centering
	\includegraphics{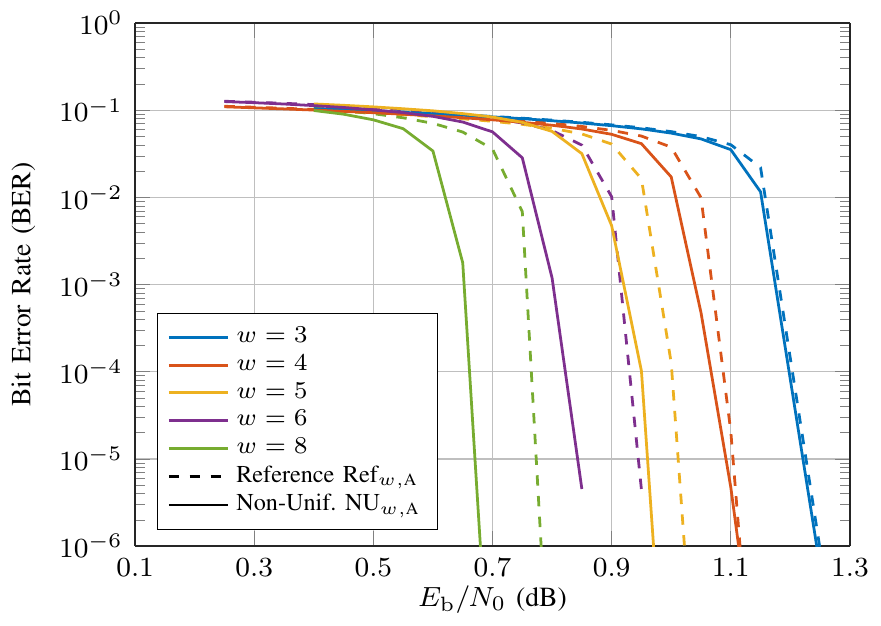}
	\vspace*{-3ex}
	\caption{Simulation results for rate $R\approx \frac{1}{2}$ codes with decoder setup~A ($M=8000$ VNs per SP)}
	\vspace*{-3ex}
	\label{fig:simres_decv1}
\end{figure}

\begin{figure}
	\centering
	\includegraphics{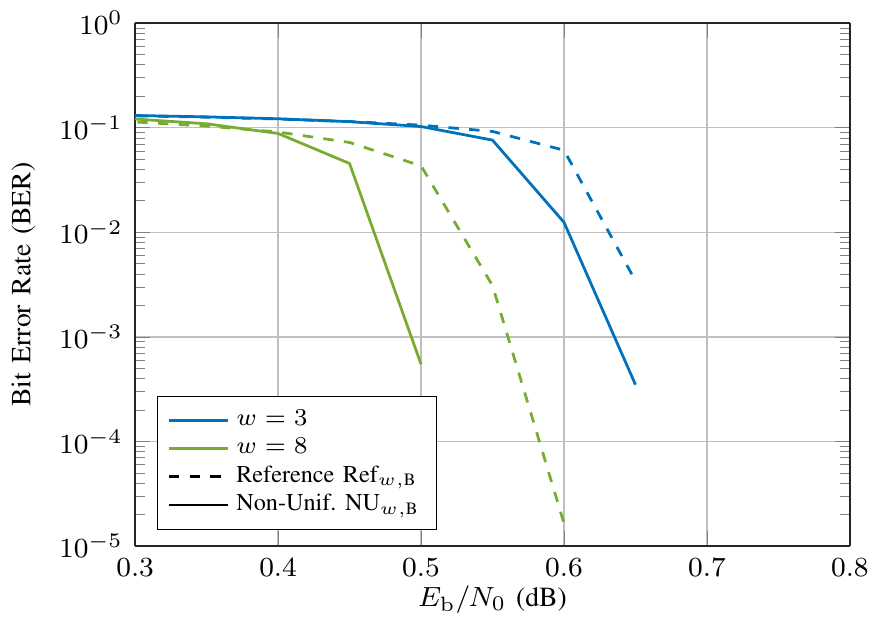}
	\vspace*{-3ex}
	\caption{Simulation results for rate $R\approx \frac{1}{2}$ codes with decoder setup~$B$ ($M=8000$ VNs per SP)}
	\vspace*{-3ex}	
	\label{fig:simres_decv2}
\end{figure}

In Fig.~\ref{fig:simres_decv1} (Decoder Setup A) and Fig.~\ref{fig:simres_decv2} (Decoder Setup B), we show the performance of the codes when simulated over a BI-AWGN channel. We use the codes from~Tab.~\ref{tab:codeselection} with $M=8000$ and both uniform and optimized non-uniform coupling. As expected, for $w=3$, the performance differs almost not at all, while for larger $w$, the performance gap becomes increasingly larger. In particular for $w=8$, we can obtain an advantage of about $0.1$\,dB and obtain codes that are quite close to capacity. Note that codes with larger $w$ also have larger decoding latency as $M$ is fixed. The same gains are observed with Decoder Setup B, where a larger baseline complexity is used during decoding. Interestingly, the rate loss of the optimized non-uniform codes is slightly larger than the rate loss of the reference, uniformly coupled, codes. This is in contrast to the results of~\cite{SchmalenISIT17} where we have shown that for $w=2$, non-uniform coupling always reduces the rate loss. However, for $w>2$, this is not necessarily the case, and we observe here an increase in the rate loss. An interesting open question is the optimization of non-uniform coupling profiles that do not lead to an additional rate loss but still improve decoding speed.\vspace*{-1ex}

\section{Conclusion}
\label{sec:conclude}
In this paper, we have considered spatially coupled LDPC codes with finite smoothing parameter. A finite smoothing parameter is important for realizing practical codes that are decoded using, e.g., windowed decoders. We have shown that by carefully optimizing the amount of coupling in each SP, we can construct codes with excellent thresholds and decoding speeds of the wave that are faster compared to conventional uniformly coupled codes. 
An interesting observation is that the dominant terms in the numerically optimized smoothing profiles $\bm{\nu}$ are only the first and the last elements while the remaining terms are negligible. The numerical simulation finally confirms the excellent, close-to-capacity performance of the constructed codes.

\end{document}